%% file: paper.tex
\documentclass[aps,prb,reprint,superscriptaddress,showpacs]{revtex4-1}
\usepackage{graphicx}
\usepackage{bm}
\usepackage{color}

\begin{document}

\title{Topological Crystalline Superconductivity in Locally Non-centrosymmetric Multilayer Superconductors}

\author{Tomohiro Yoshida}
\affiliation{Department of Physics, Niigata University, Niigata 950-2181, Japan}
\author{Manfred Sigrist}
\affiliation{Theoretische Physik, ETH-Z\"urich, 8093 Z\"urich, Switzerland}
\author{Youichi Yanase}
\affiliation{Department of Physics, Niigata University, Niigata 950-2181, Japan}

\date{\today}

\renewcommand{\k}{{\bm k}}

\begin{abstract}
Topological crystalline superconductivity in the locally non-centrosymmetric multilayer superconductors (SCs) is proposed. 
We study the odd-parity pair-density wave (PDW) state induced by the spin-singlet pairing interaction 
through the spin-orbit coupling. 
It is shown that the PDW state is a topological crystalline SC protected by a mirror symmetry,  
although it is topologically trivial according to the classification based on the standard topological periodic table. 
The topological property of the mirror subsectors is intuitively explained by adiabatically changing the BdG Hamiltonian. 
A subsector of the bilayer PDW state reduces to the two-dimensional non-centrosymmetric SC, while 
a subsector of trilayer PDW state is topologically equivalent to the spinless $p$-wave SC. 
Chiral Majorana edge modes in trilayers can be realized without Cooper pairs in the spin-triplet channel 
and chemical potential tuning.
\end{abstract}


\maketitle

Topologically nontrivial phases of superconductors (SCs) have evolved into one of the major research topics
of modern condensed matter physics recently~\cite{PRB.61.10267,PRL.86.268,PU.44.131,PRB.78.195125,PRL.100.096407,PRB.79.060505,PRL.102.187001,PRB.79.094504,PRL.103.020401,PRB.82.134521,PRB.81.220504,PRL.104.040502,*PRB.81.125318,*PRL.105.077001,PRL.105.097001,RMP.83.1057,JPSJ.81.011013,Fu2014}. 
A characteristic feature of topological SCs is the fully gapped bulk spectrum accompanied by 
topologically protected gapless edge states.
Many of the topological superconducting states are realized in odd-parity SCs,
and one of the most extensively studied examples is the chiral $p_x \pm ip_y$-wave SC~\cite{PRB.61.10267,PRL.86.268}. 
However, only few materials are considered as possible hosts of odd-parity superconductivity, 
because the conditions for spin-triplet pairing are quite unfavorable in most cases.
So far, Sr$_2$RuO$_4$~\cite{JPSJ.81.011009} and some uranium-based heavy fermion compounds~\cite{RMP.74.235,JPSJ.81.011003} 
show strong evidence for the spin-triplet odd-parity superconductivity,
but unfortunately their superconducting gap might have nodes on the Fermi surface.
Recently, odd-parity topological superconductivity in a doped topological insulator Cu$_x$Bi$_2$Se$_3$ 
has been proposed~\cite{PRL.105.097001,Fu2014}, however, experimental results are under debate~\cite{PRL.107.217001,PRL.110.117001}.

\begin{figure}[tbp]
  \includegraphics[width=85mm]{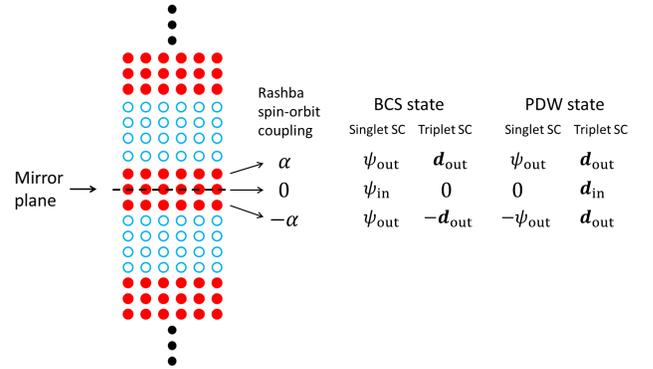}
  \caption{(Color online) Schematic figure of trilayer system.
  The filled (open) circles represent the 2D superconducting (normal spacer) layers.
  The dashed line denotes the mirror plane. 
  Attached lists provide information on the
  layer-dependent Rashba spin-orbit coupling and the order parameters in the BCS and PDW states.}
  \label{fig1}
\end{figure}
In a recent study we showed that odd-parity superconductivity occurs naturally in multilayer systems 
with layer-dependent spin-orbit coupling arising from the local lack of inversion symmetry~\cite{PRB.86.134514}. 
We will consider here such locally non-centrosymmetric systems composed of the blocks of superconducting layers, 
e.g. trilayer systems as depicted in Fig.~\ref{fig1}. 
Here the layer-dependent Rashba spin-orbit coupling is responsible for unusual electronic and superconducting 
properties~\cite{JPSJ.81.034702}. 
The coupling constant of the Rashba spin-orbit coupling shows the layer-dependence, 
$(\alpha_1,\alpha_2,\alpha_3)=(\alpha,0,-\alpha)$, ensured by the global inversion symmetry. 
We have shown that in such a system an odd-parity superconducting state can be stabilized by a magnetic field, 
even if the zero-field phase is the even-parity state (see Fig.~\ref{fig1})~\cite{PRB.86.134514}. 
To be precise, the order parameter in the spin-singlet channel changes sign between the outer-most layers in the 
field-induced superconducting state (see Fig.~\ref{fig1}). 
Considering the spatially modulating order parameter in the trilayer, 
we call it the ``pair-density wave (PDW) state''~\cite{PRL.102.207004}. 
Multilayer structures of this kind are not only theoretical constructs, but have indeed been produced recently, 
for example, in the artificially grown superlattices CeCoIn$_5$/YbCoIn$_5$~\cite{NP.7.849,PRL.109.157006,PRL.112.156404} 
and in transition-metal-oxide interfaces~\cite{PRB.88.241104}.
The PDW state is stabilized when the three conditions, (a) Pauli-limited SC, (b) quasi-two-dimensional structure, 
and (c) large spin-orbit coupling, are satisfied. These conditions are naturally satisfied in the heavy fermion 
superlattice CeCoIn$_5$/YbCoIn$_5$~\cite{NP.7.849,PRL.109.157006,PRL.112.156404}. 
Furthermore, the recent technology enabled the artificial tuning of the superlattice structure~\cite{PRL.112.156404}. 
Thus, we may expect that the PDW state can be stabilized in a superlattice  CeCoIn$_5$/YbCoIn$_5$, 
although no experimental evidence has been reported so far. 
In view of the experimental and theoretical status, the discussion of topological features 
of the PDW state is well motivated. 

Topological aspects of bilayer PDW state {\it in the absence of} a magnetic field have been investigated 
by Nakosai {\it et al}.~\cite{PRL.108.147003}. 
They showed that the bilayer PDW state is a topological state protected by a Z$_2$ invariant 
when (and only when) the Fermi level lies in the hybridization gap between the bonding and anti-bonding bands. 
The field-induced PDW phase in the multilayer system has not been investigated in this respect so far.

First, we consider the topological properties of the PDW state on the basis of the 
{\it so-called} topological periodic table~\cite{PRB.78.195125}.
When time-reversal symmetry is broken by a magnetic field, the symmetry class of the state is D. 
The two-dimensional (2D) system in the class D is characterized by an integer topological number, 
the Chern number~\cite{PRL.49.405,AP.160.343}. 
However, the Chern number must be zero in the time-reversal invariant system, and the magnetic field does not change 
the Chern number without closing the gap. 
According to the numerical analysis of the Bogoliubov-de-Gennes (BdG) equation, 
the magnetic field does not close the gap in the PDW state~\cite{PRB.86.134514}. 
Thus, this shows that the field-induced PDW state is topologically trivial in terms of the classification based on 
the topological periodic table.

On the other hand, recent developments in the classification scheme of topological phases shed new light on 
topological phases protected by the crystal symmetry~\cite{PRL.106.106802,PE.55.20,PRL.111.056403,PRB.88.075142,PRL.111.087002,PRB.88.125129,JPSJ.82.113707,Shiozaki-Sato,Chiu-Schnyder}. 
The ''topological crystalline SCs'' have been classified relying on 
the mirror, inversion, rotation and magnetic point group symmetry~\cite{PRB.88.075142,PRB.88.125129,Shiozaki-Sato}. 
The spin-triplet superconducting/superfluid states in Sr$_2$RuO$_4$~\cite{PRL.111.087002}, UPt$_3$~\cite{JPSJ.82.113707}, 
and $^3$He~\cite{PE.55.20} have been discussed from this point of view.  
In this letter, we will show that the spin-singlet PDW state in trilayers is generally a topological crystalline 
SC protected by the mirror symmetry. This is, to our knowledge, the first proposal for the topological crystalline SC 
without requiring the pairing interaction in the spin-triplet channel.

We consider the mean-field BdG Hamiltonian for the 2D multilayer SC, 
\begin{eqnarray}
  {\cal H}&=& \sum_{{\bm k},s,s',m} [\xi({\bm k}) \sigma_0 
+ \alpha_m {\bm g}({\bm k})\cdot{\bm \sigma} 
- \mu_{\rm B}H \sigma_z]_{ss'}
c^\dagger_{{\bm k}sm}c_{{\bm k}s'm} \nonumber \\
  &&+t_\perp\sum_{{\bm k},s,\langle m,m'\rangle}c^\dagger_{{\bm k}sm}c_{{\bm k}sm'} \nonumber \\
  &&+\frac{1}{2}\sum_{{\bm k},s,s',m}[\Delta_{ss'm}({\bm k})c^\dagger_{{\bm k}sm}c^\dagger_{-{\bm k}s'm}+{\rm h.c.}],
  \label{eq1}
\end{eqnarray}
where ${\bm k}$, $s$, and $m$ $(=1, \dots, M) $ are indices of momentum, spin, and layer, respectively. 
We assume the simple dispersions $\xi({\bm k})=-2t(\cos k_x+\cos k_y)-\mu$ and ${\bm g}({\bm k})=(-\sin k_y,\sin k_x,0)$. 
The latter describes the Rashba spin-orbit coupling, whereby the coupling constant $\alpha_m$ is layer-dependent.  
Nearest-neighbor layers are coupled by the hopping matrix element $t_\perp$. 
We focus on the intra-layer Cooper pairing which is relevant for 2D SCs, as realized in
CeCoIn$_5$/YbCoIn$_5$ superlattices~\cite{NP.7.849,PRL.109.157006,PRL.112.156404} and $\delta$-doped SrTiO$_3$~\cite{PRB.88.241104}, 
although an inter-layer Cooper pairing has been considered for Cu$_x$Bi$_2$Se$_3$~\cite{PRL.105.097001,Fu2014}.
The layer-dependent order parameter can then be parameterized by 
$\hat{\Delta}_m({\bm k})=[\psi_m({\bm k})+{\bm d}_m({\bm k})\cdot{\bm \sigma}]i\sigma_y$,
where $\psi_m({\bm k})$ and ${\bm d}_m({\bm k})$ represent the spin-singlet and spin-triplet components of order parameters 
on the layer $m$, respectively. 
For simplicity, we assume the $S$+$p$-wave pairing state, in which the dominant $s$-wave order parameter $\psi_m({\bm k})=\psi_m$ 
is mixed with the spin-triplet $p$-wave component through spin-orbit coupling and pairing interaction. The latter has the structure 
${\bm d}_m({\bm k})=a_m(-\sin k_y,\sin k_x,0)+ib_m(\sin k_x,\sin k_y,0)$, obtained  
by solving the BdG equation~\cite{JPSJ.83.013703}. 
In the following we analyze the two competing solutions of the BdG equation: (1) the ''BCS state'' 
with $\psi_m(\k) = \psi_{M+1-m}(\k)$ and ${\bm d}_m(\k) = - {\bm d}_{M+1-m}(\k)$
and (2) the ''PDW state'' 
where $\psi_m(\k) = -\psi_{M+1-m}(\k)$ and ${\bm d}_m(\k) = {\bm d}_{M+1-m}(\k)$. 
We now assume a pairing mechanism favoring spin-singlet pairing, as often given by electron-phonon coupling or  
antiferromagnetic spin fluctuation. Thus, 
the BCS state is stabilized by the inter-layer Josephson coupling at zero magnetic field. 
However, the PDW state is stabilized by spin-orbit coupling 
in the high magnetic field region at sufficiently low temperatures~\cite{PRB.86.134514}.

Now we define the topological invariant of multilayer SCs protected by the mirror symmetry, by means of the mirror Chern number. 
The BdG Hamiltonian is represented as, $  {\cal H}=\frac{1}{2}\sum_{{\bm k}}\Psi^\dagger_{\bm k}{\cal H}({\bm k})\Psi_{\bm k} $ 
with use of Nambu operators $\Psi^\dagger_{\bm k}=(c^\dagger_{{\bm k}sm},c_{-{\bm k}sm})$ in $ 4 \times M $ dimension. 
The mirror symmetry with respect to the central $xy$-plane is obeyed, 
\begin{eqnarray}
  {\cal M}_{xy}^\pm{\cal H}({\bm k}){\cal M}_{xy}^{\pm\dagger}={\cal H}({\bm k}).
  \label{eq2}
\end{eqnarray}
${\cal M}^\pm_{xy}$ is the mirror reflection operator in the particle-hole space (see Appendix A). 
We introduce ${\cal M}_{xy}^{+}$ for the BCS state and ${\cal M}_{xy}^{-}$ for the PDW state, respectively.  
Equation (\ref{eq2}) guarantees that the BdG Hamiltonian can be block-diagonalized in the eigenbasis
of ${\cal M}_{xy}^\pm$. Thus, the system is divided into the two subsectors corresponding to the 
block Hamiltonian ${\cal H}_\lambda^\pm({\bm k})$ with $ \lambda=\pm i $ as eigenvalues of ${\cal M}_{xy}^\pm$. 
We now define the mirror Chern number $\nu(\lambda)$, as the Chern number 
of the subsector Hamiltonian~\cite{PE.55.20,PRL.111.087002} (see Appendix B).
The topological protection of the mirror Chern number is guaranteed in some topological classes 
characterized by the symmetries of subsector Hamiltonian $ {\cal H}_\lambda^\pm({\bm k}) $~\cite{PRB.78.195125}. 
Important here are the time-reversal, particle-hole and chiral symmetry (see Appendix C).

For illustration we first discuss the bilayer system.
We obtain the subsector Hamiltonian for the $\lambda=i$ sector as 
\begin{eqnarray}
  {\cal H}_{\lambda= i}^\pm({\bm k})=\left(
  \begin{array}{cc}
    {\cal H}'({\bm k}) + t_\perp\sigma_z & \pm i[\psi-{\bm d}({\bm k})\cdot{\bm \sigma}]\sigma_y \\
    \mp i\sigma_y[\psi^\ast-{\bm d}^\ast({\bm k})\cdot{\bm \sigma}] & -{\cal H}^{'T}(-{\bm k})  \pm t_\perp\sigma_z
  \end{array}
  \right), \nonumber \\
  \label{eq3}
\end{eqnarray}
where ${\cal H}'({\bm k})=\xi({\bm k})\sigma_0-\mu_{\rm B}H\sigma_z-\alpha{\bm g}({\bm k})\cdot {\bm \sigma}$.
The subsector Hamiltonian for $\lambda=-i$ is obtained by changing the sign of $t_\perp$, as $t_\perp\rightarrow -t_\perp$. 
For the BCS state,
although the particle-hole symmetry in the original BdG Hamiltonian is conserved,
we cannot rely on this symmetry in the subsector Hamiltonian  
unless the special condition ${\cal H}_{\lambda=i}^+({\bm k})={\cal H}_{\lambda=-i}^+({\bm k})$, namely $t_\perp =0$, 
is satisfied (demonstrated in the supplementary material). On the other hand, the chiral symmetry is conserved in this subsector at $H=0$. 
Therefore, in the absence of a magnetic field the symmetry class is AIII which is topologically trivial 
in 2D~\cite{PRB.78.195125}. If the chiral symmetry is broken by a magnetic field, both subsectors 
belong to the class A, which is characterized by an integer topological invariant~\cite{PRB.78.195125}.
However, both subsectors are topologically trivial, $\nu(\lambda) = 0$, or the gap is closed 
under the realistic condition, $|\psi| \ll t_\perp $.

For the odd-parity PDW state, time-reversal symmetry in the subsector Hamiltonian is ill-defined for $t_\perp \neq 0$, while
the particle-hole symmetry is conserved.  
Thus, the subsector belongs to the symmetry class D unless $(t_\perp, H)=(0,0)$. 
Interestingly, each subsector is equivalent to the BdG Hamiltonian of a 2D non-centrosymmetric superconductor (NCSC)~\cite{NCSC} 
with the fictitious magnetic field $\mu_{\rm B}H\pm t_\perp$, 
whose topological property has already been clarified~\cite{PRB.79.060505,PRB.79.094504,PRL.103.020401,PRB.82.134521,PRL.104.040502,*PRB.81.125318,*PRL.105.077001}. 
The dominantly spin-singlet pairing state $|{\bm d}({\bm k})| < |\psi|$ can be topologically nontrivial, 
when the effective magnetic field $\mu_{\rm B}H \pm t_\perp$ satisfies the condition 
$\sqrt{(4t+\mu)^2+|\psi|^2}< \left| \mu_{\rm B}H\pm t_\perp \right| <\sqrt{(4t-\mu)^2+|\psi|^2}$,  
[$\sqrt{\mu^2+|\psi|^2}< \left| \mu_{\rm B}H\pm t_\perp \right| <\sqrt{(4t-\mu)^2+|\psi|^2}$]
for $\mu\leq -2t$ [$-2t<\mu\leq 0$]~\cite{PRB.82.134521}. 
Although great effort has been devoted to the realization of this condition in semiconductor devices~\cite{Mourik}, 
this condition needs fine tuning of the chemical potential and is rather unrealistic in metals.

For $H=0$, this condition is indeed equivalent to the criterion for a $Z_2$ topological SC 
without relying on the mirror symmetry~\cite{PRL.108.147003}. 
This means that the nontrivial $Z_2$ topological number in the original BdG Hamiltonian (class DIII) 
is obtained by the mirror Chern number of the subsectors (class D). 
This is analogous to the fact that some $Z_2$ topological insulators are characterized 
by the spin Chern number~\cite{PRL.97.036808}.
Our analysis sheds light on the analogy between the 2D NCSC and the $Z_2$ nontrivial bilayer SC,
the former being equivalent to a mirror subsector of the latter. The interlayer coupling $t_\perp$ plays the same role 
as the magnetic field in the former. 
Although the $Z_2$ number of the original BdG Hamiltonian is not a topological invariant in the presence of the magnetic field, 
the mirror Chern number is topologically protected. 
Therefore, the mirror Chern number is useful to indicate the topological property of field-induced superconducting states.

\begin{figure}[tbp]
  \begin{center}
    \begin{tabular}{cc}
      \begin{minipage}{0.5\hsize}
        \includegraphics[width=42mm]{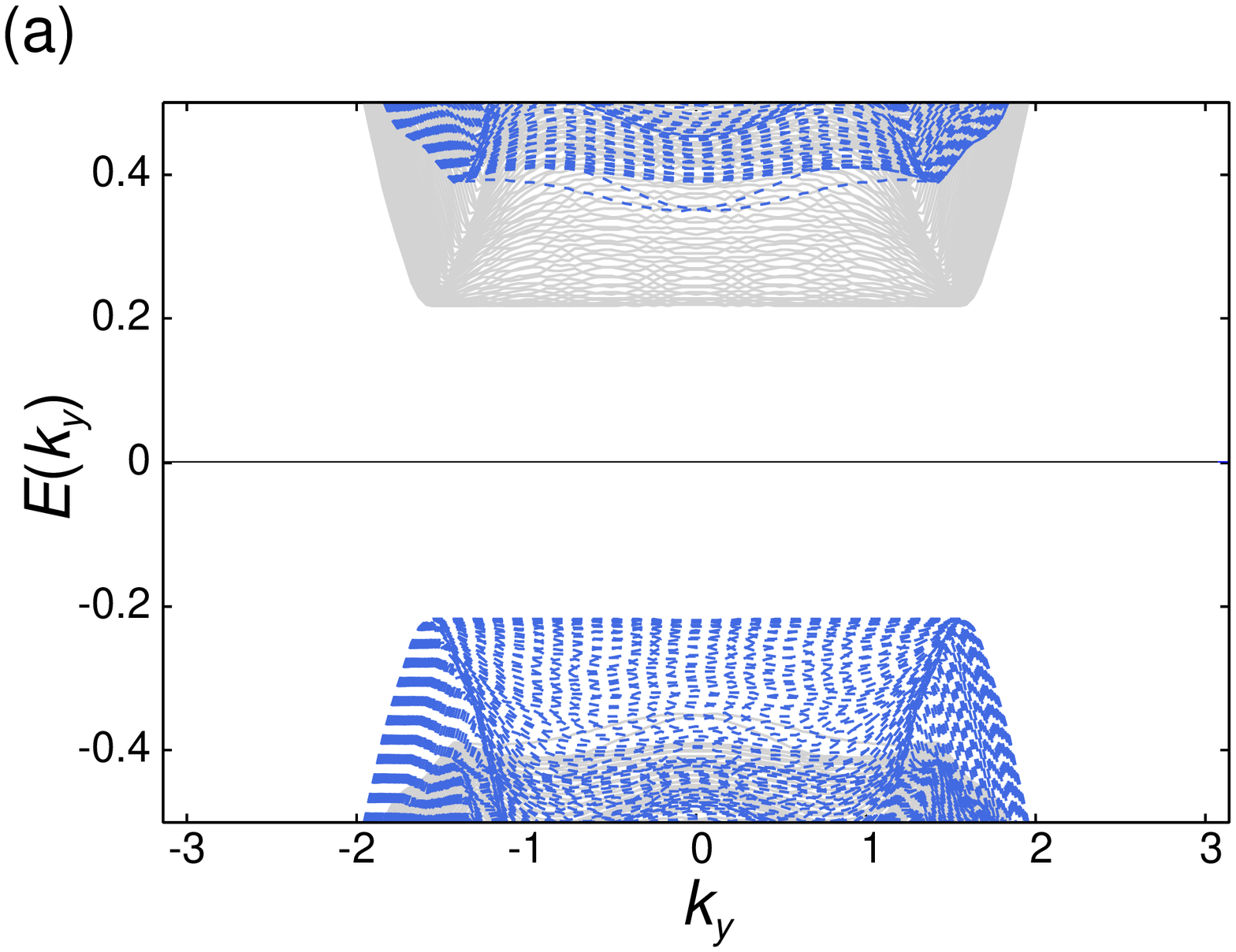}
      \end{minipage}
      \begin{minipage}{0.5\hsize}
        \includegraphics[width=42mm]{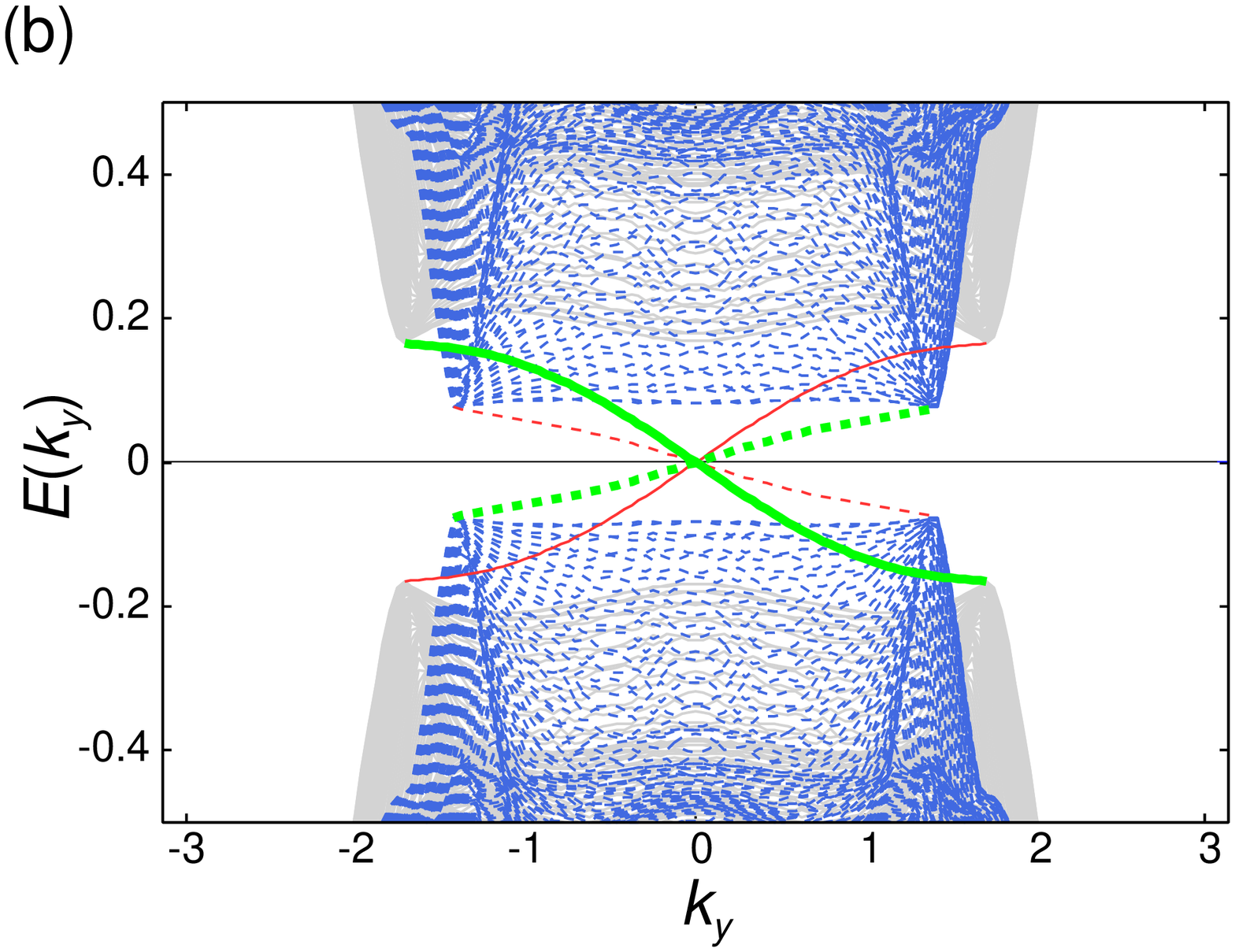}
      \end{minipage}
    \end{tabular}
    \begin{tabular}{cc}
      \begin{minipage}{0.5\hsize}
        \includegraphics[width=42mm]{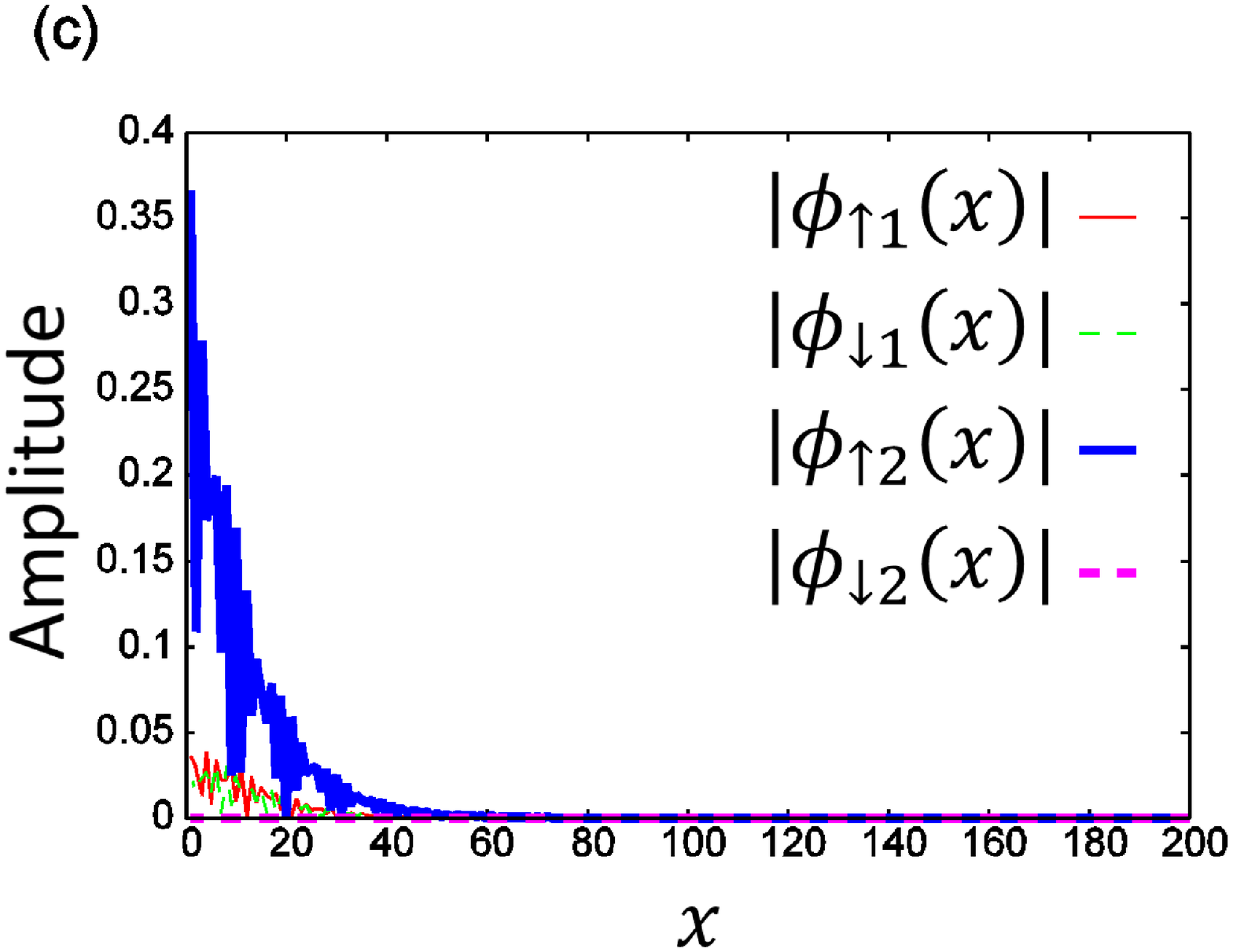}
      \end{minipage}
      \begin{minipage}{0.5\hsize}
        \includegraphics[width=42mm]{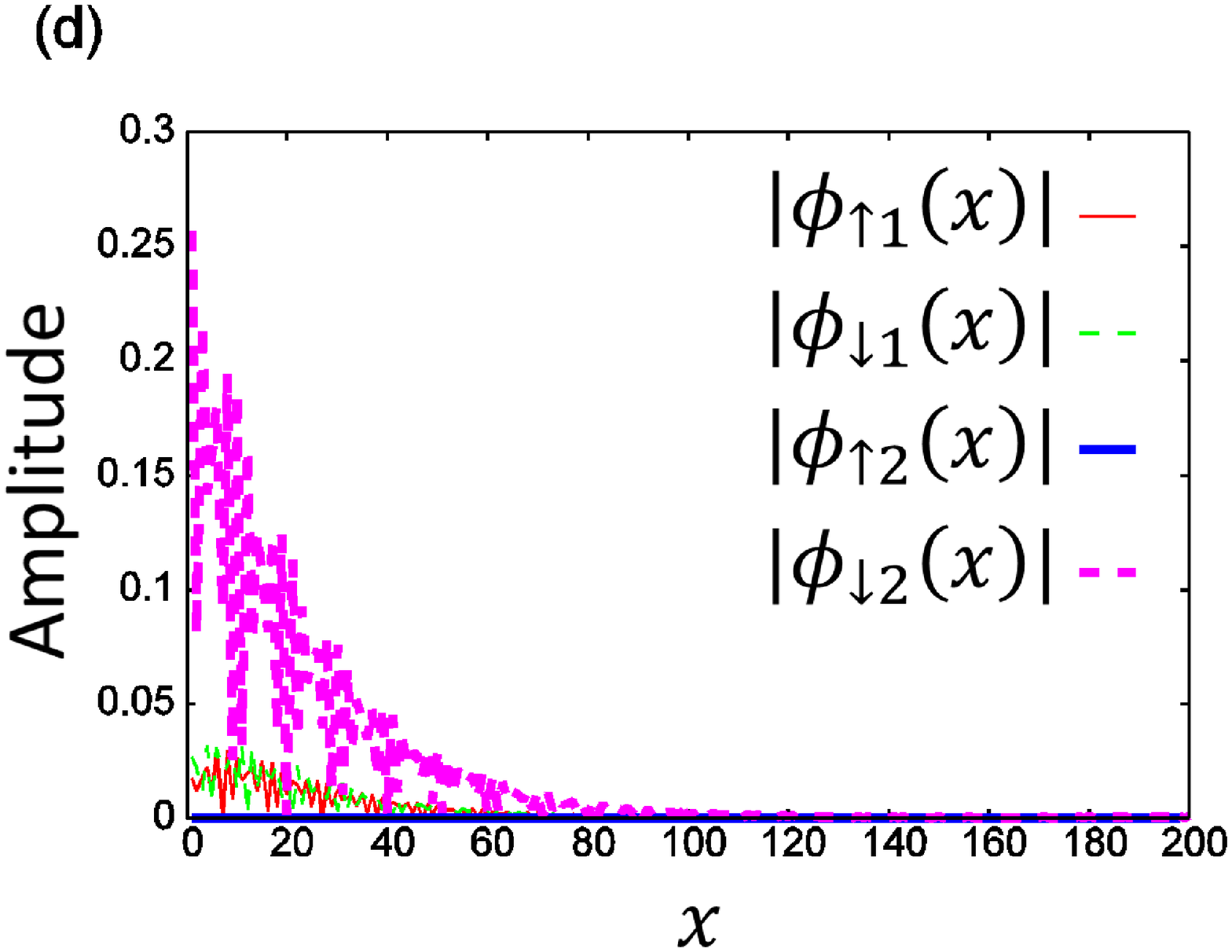}
      \end{minipage}
    \end{tabular}
    \caption{(Color online) Energy spectra of (a) the BCS state and (b) the PDW state with open boundaries at $x=1$ and $x=200$. 
      The solid and dashed lines in (b) show the Majorana edge modes in $\lambda=i$ and $\lambda=-i$ subsectors, 
      respectively.
      Thick (green) lines show the edge states near the boundary $x=1$, 
      while thin (red) lines show the edge states near $x=200$. 
      We take $t=1$, $\mu=-2$, $\mu_{\rm B}H=0.3$, $\alpha=0.3$, $t_\perp=0.1$, $\psi_{\rm out}=\psi_{\rm in}=0.5$, 
      $a_{\rm out}=a_{\rm in}=-0.05$, and $b_{\rm out}=b_{\rm in}=0.1$.
      (c) and (d) illustrate the wave function of Majorana modes localized around $x=1$. 
      Amplitude of spin- and layer-resolved wave function, $\phi_{sm}(x) = \langle x, sm | E=0 \rangle$, is shown. 
      The Majorana state resides dominantly on the center layer ($m=2$) with up spin for the subsector $ \lambda = i $ (c) and with down spin for  $ \lambda = -i $ (d). 
    }
    \label{fig2}
  \end{center}
\end{figure}

We now turn to the trilayer system to show the most important results of this paper. 
We consider the trilayer structure conserving the mirror symmetry (see Fig.~\ref{fig1}), and 
adopt the layer-dependent Rashba spin-orbit coupling $(\alpha_1,\alpha_2,\alpha_3)=(\alpha,0,-\alpha)$.
The layer-dependent order parameters are shown in Fig.~\ref{fig1}.
Using the mirror operator with respect to the central $xy$-plane, 
the BdG Hamiltonian is again block-diagonalized into the mirror subsectors.
We show the subsector Hamiltonian for the BCS state in Appendix B.
The subsector belongs to the class A for $ H \neq 0 $ and to the class AIII for $H=0$, if $t_\perp \ne 0$. 
We confirmed that the mirror Chern number is zero or the gap is closed as in bilayers. 
Thus, topological superconductivity is not realized in the BCS state. 
Indeed, Fig.~\ref{fig2}(a) shows no zero energy Majorana mode, indicating the topologically trivial property.

In contrast, the PDW phase represents a topological crystalline superconducting state.
We obtain the subsector Hamiltonian
\begin{widetext}
  \begin{eqnarray}
    {\cal H}_{\lambda=i}^-({\bm k})=\left(
    \begin{array}{cccccc}
      \xi_\uparrow({\bm k}) & \alpha k_+ & \sqrt{2}t_\perp & 0 & -d_{{\rm out}-}({\bm k}) & -\psi_{\rm out} \\
      \alpha k_- & \xi_\downarrow({\bm k}) & 0 & 0 & \psi_{\rm out} & d_{{\rm out}+}({\bm k}) \\
      \sqrt{2}t_\perp & 0 & \xi_\uparrow({\bm k}) & -d_{{\rm in}-}({\bm k}) & 0 & 0 \\
      0 & 0 & -d_{{\rm in}-}^\ast({\bm k}) & -\xi_\uparrow({\bm k}) & -\sqrt{2}t_\perp & 0 \\
      -d_{{\rm out}-}^\ast({\bm k}) & \psi_{\rm out}^\ast & 0 & -\sqrt{2}t_\perp & -\xi_\uparrow({\bm k}) & \alpha k_- \\
      -\psi_{\rm out}^\ast & d_{{\rm out}+}^\ast({\bm k}) & 0 & 0 & \alpha k_+ & -\xi_\downarrow({\bm k})
    \end{array}
    \right),  
    \label{eq4}
  \end{eqnarray}
\end{widetext}
for $\lambda = i$. We denote $\xi_s({\bm k})=\xi({\bm k})-(\sigma_z)_{ss}\mu_{\rm B}H$, $k_\pm=\sin k_y\pm i\sin k_x$, and 
  $d_{{\rm out} ({\rm in})\pm}({\bm k})=d_{{\rm out}({\rm in})}^{(x)}({\bm k})\pm id_{{\rm out}({\rm in})}^{(y)}({\bm k})$.
The subsector Hamiltonian for $\lambda=-i$ is shown in the supplemental material.
Both subsectors belong to the symmetry class D independent of the magnetic field,
if $t_\perp \ne 0$. 
Therefore, the mirror Chern number is a topological invariant. We obtain a nontrivial mirror Chern number 
$\nu(\lambda=\pm i) = \mp 1$, almost independent of the parameters. 
In contrast to the bilayer PDW state, this topologically nontrivial superconducting state is realized 
without having to rely on a special choice of parameters. 
Because the mirror Chern number is odd, the trilayer PDW state is also a $Z_2$ topological 
superconducting state at $H = 0$, although the magnetic field is required for the thermodynamic stability of 
the PDW state~\cite{PRB.86.134514}.

An intuitive understanding of our result can be obtained by adiabatically deforming the subsector Hamiltonian 
$ {\cal H}_{\lambda}^-({\bm k})$.  
The interlayer coupling $t_\perp$ is decreased to zero without closing the gap as long as 
the spin-triplet component ${\bm d}_{{\rm in}}({\bm k})$ is finite. 
The topology does not change through this adiabatic deforming. 
Then, the finite mirror Chern number originates from the decoupled $2 \times 2$ matrix 
in the center of $6 \times 6$ matrix of Eq.~(\ref{eq4}), which denotes a spinless chiral $p$-wave SC. 
It has been shown that the spinless chiral p-wave SC is topologically nontrivial~\cite{PRB.61.10267} 
and the Chern number is $\pm 1$ (see Appendix D).
Indeed, we obtained the nontrivial mirror Chern number $\nu(\pm i) = \mp 1$, 
which is identified as the Chern number originates from the decoupled $2 \times 2$ matrix 
in the limit $t_{\perp} \rightarrow 0$. 
Now it became apparent that no fine tuning of the chemical potential is needed.
The other $4 \times 4$ matrix decoupled in the subsector Hamiltonian describes the 2D Rashba-type NCSC which has been proposed to be 
a topological $s$-wave SC~\cite{PRL.103.020401,PRB.82.134521,PRL.104.040502,*PRB.81.125318,*PRL.105.077001}. 
However, we do not assume a fine tuning of the chemical potential which is required in their proposals.

We emphasize that the Cooper pairing in the $p$-wave channel ${\bm d}_{{\rm in/out}}({\bm k})$ is not needed 
for the topological crystalline superconductivity, although it played an important role in the above 
intuitive explanation. This is understood from the fact that ${\bm d}_{\rm in/out}({\bm k})$ is decreased to zero 
without closing the gap when the interlayer hopping $t_\perp$ is finite~\cite{PRB.86.134514}. 
Thus, the topology is equivalent between the Hamiltonian 
for $t_\perp=0$ and ${\bm d}_{\rm in/out}({\bm k})\neq 0$ (as in the above intuitive explanation) 
and that for $t_\perp \neq 0$ and ${\bm d}_{\rm in/out}({\bm k})=0$ (as we consider here). 
This means that the topological crystalline superconductivity is realized without any attractive interaction 
in the spin-triplet channel. Once the PDW state is stabilized in the trilayer system, it is a topological crystalline SC. 

In order to verify the bulk-edge correspondence, we show the presence of edge states in the trilayer SCs. 
Figures \ref{fig2}(a) and (b) show the energy spectra of BCS state and PDW state, respectively, 
for a ribbon-shaped system with open boundaries along $x$-axis and translational invariance along $y$-direction. 
Consistent with the vanishing mirror Chern number, no subgap edge state appear in the BCS state.
In contrast, we find two chiral Majorana edge modes in the PDW state. One comes from the $\lambda = i$ subsector
(solid lines) and the other comes from the $\lambda = - i$ subsector (dashed lines). 
These modes are not Kramers pairs, because the time-reversal symmetry is broken by the magnetic field. 
We confirmed that the presence of these Majorana modes is robust against the change of parameters, 
such as variations of $\psi_{{\rm in/out}}$, $a_{{\rm in/out}}$, $b_{{\rm in/out}}$, $t_\perp$, $\alpha$, and $\mu$.

In Figs.~\ref{fig2}(c) and (d), we show the spatial profiles of the zero-energy Majorana modes localized around the edge. 
Large probability density on the inner layer $|\phi_{s2}(x)| = |\langle x, s2 | E=0 \rangle|$ is also shown. This means that 
the Majorana state mainly originates from the inner layer, as expected from the intuitive explanation discussed above.

In this letter we have focused on the 2D multilayer SCs, but the topologically nontrivial properties also appear 
in the three-dimensional (3D) system. When we take into account an inter-multilayer coupling through normal spacer layers 
(see Fig.~\ref{fig1}) and consider the 3D Brillouin zone, the BdG Hamiltonian conserves the mirror reflection symmetry as 
${\cal M}_{xy}^\pm{\cal H}(k_x,k_y,k_z){\cal M}_{xy}^\pm={\cal H}(k_x,k_y,-k_z)$. 
Thus, the mirror symmetry defined by Eq.~(\ref{eq2}) is satisfied in the mirror invariant planes, $k_z=0$ and $\pi$.
We can define the mirror Chern number in these 2D mirror invariant planes, and we indeed obtain 
a nontrivial mirror Chern number at both $k_z=0$ and $\pi$ for a small inter-multilayer coupling. 
We confirmed that Majorana cones appear on [100] and [010] surfaces where the mirror symmetry is conserved.

Analyzing topological properties of multilayer SCs we found that the PDW state is
a topological crystalline superconducting phase protected by the mirror symmetry. 
We stress that 
a purely $s$-wave PDW state in trilayers can be a topological SC accompanied by the Majorana fermion on its edge
without the tuning of chemical potential, which is necessary in the bilayer PDW state~\cite{PRL.108.147003} and the 1D and 2D NCSC~\cite{PRL.103.020401,PRL.104.040502,*PRB.81.125318,*PRL.105.077001}. 
This finding significantly expands the possibility of realizing the topological SC because most SCs have a $s$-wave symmetry. 
It is straightforward to extend our analysis to more than three layers and we find that the PDW state is a topological SC independent of parameters, if the number of layers is odd. 
Thus, the design of the topological crystalline SC is feasible for artificially grown multilayers 
using the available technology~\cite{NP.7.849,PRL.109.157006,PRL.112.156404,PRB.88.241104}. 
The superlattice CeCoIn$_5$/YbCoIn$_5$~\cite{NP.7.849} is considered to be a $D$+$p$-wave SC, 
and will be similarly a topological crystalline SC as will be discussed elsewhere~\cite{Yoshida-Sigrist-Yanase}. 

The authors are grateful to D.~Maruyama, Y.~Matsuda, T. Morimoto, T.~Shibauchi, M.~Shimozawa, K. Shiozaki, 
A.~P.~Schnyder, Y.~Ueno, and A.~Yamakage for fruitful discussions. 
T.~Y. is supported by a JSPS Fellowship for Young Scientists. 
This work was supported by KAKENHI Grants Numbers 24740230, 25103711, and 15K05164.

\appendix{
  \section{MIRROR OPERATOR FOR MULTILAYER SUPERCONDUCTOR}

In this section, we introduce the mirror operator for multilayer SCs. 
For the clarity of discussions, we describe the BdG Hamiltonian with use of 
the normal state Hamiltonian ${\cal H}_0({\bm k})$ and the gap function $\Delta({\bm k})$ as, 
\begin{eqnarray}
  {\cal H}({\bm k})=\left(
  \begin{array}{cc}
    {\cal H}_0({\bm k}) & \Delta({\bm k}) \\
    \Delta^\dagger({\bm k}) & -{\cal H}_0^T(-{\bm k})
  \end{array}
  \right). 
  \label{SM_eq1}
\end{eqnarray} 
Because the mirror symmetry with respect to the $xy$-plane is evidently conserved in the multilayer systems, 
the normal state Hamiltonian ${\cal H}_0({\bm k})$ is invariant for the mirror 
reflection operator ${\cal M}_{xy}$ as, 
\begin{eqnarray}
  {\cal M}_{xy}{\cal H}_0({\bm k}){\cal M}^\dagger_{xy}={\cal H}_0({\bm k}). 
  \label{SM_eq2}
\end{eqnarray}
The mirror reflection operator transforms the momentum ${\bm k}$ as, 
\begin{eqnarray}
  {\bm k}=(k_x,k_y,k_z)\rightarrow (k_x,k_y,-k_z), 
  \label{SM_eq3}
\end{eqnarray}
and the spin as, 
\begin{eqnarray}
  {\bm s}=(s_x,s_y,s_z)\rightarrow(-s_x,-s_y,s_z), 
  \label{SM_eq4}
\end{eqnarray}
respectively. 
As we focus on the 2D system, the momentum is invariant under the mirror reflection, 
while the spin is subject to the $\pi$-rotation around the $z$ axis. 
In addition to these fundamental degrees of freedom, the multilayer systems also have the index for layer $m$. 
By taking into account the reversal of layers, the mirror reflection operator for the normal part 
is given by 
\begin{eqnarray}
  \hspace{-8mm}
  {\cal M}_{xy}=\left(
  \begin{array}{cc}
    i\sigma_z & 0 \\
    0 & i\sigma_z
  \end{array}
  \right)\times\left(
  \begin{array}{cc}
    0 & \sigma_0 \\
    \sigma_0 & 0 
  \end{array}
  \right)=\left(
  \begin{array}{cc}
    0 & i\sigma_z \\
    i\sigma_z & 0
  \end{array}
  \right), 
  \label{SM_eq5}
\end{eqnarray}
for bilayers, while it is given by
\begin{eqnarray}
  {\cal M}_{xy}&=&\left(
  \begin{array}{ccc}
    i\sigma_z & 0 & 0 \\
    0 & i\sigma_z & 0 \\
    0 & 0 & i\sigma_z 
  \end{array}
  \right)\times\left(
  \begin{array}{ccc}
    0 & 0 & \sigma_0 \\
    0 & \sigma_0 & 0 \\
    \sigma_0 & 0 & 0
  \end{array}
  \right) \nonumber \\
 &=&\left(
  \begin{array}{ccc}
    0 & 0 & i\sigma_z \\
    0 & i\sigma_z & 0 \\
    i\sigma_z & 0 & 0
  \end{array}
  \right),
  \label{SM_eq6}
\end{eqnarray}
for trilayers. 
It is straightforward to obtain the mirror reflection operator for more than three layers. 
We confirmed that the normal state Hamiltonian ${\cal H}_0({\bm k})$ is invariant for these operators.

When the gap function has a well-defined mirror-parity 
as ${\cal M}_{xy}\Delta({\bm k}){\cal M}_{xy}^T=\pm\Delta({\bm k})$, 
the BdG Hamiltonian ${\cal H}({\bm k})$ is invariant under the mirror reflection as, 
\begin{eqnarray}
  {\cal M}_{xy}^\pm{\cal H}({\bm k}){\cal M}_{xy}^{\pm\dagger}={\cal H}({\bm k}),
  \label{SM_eq7}
\end{eqnarray}
where the mirror operator in the particle-hole space ${\cal M}_{xy}^\pm$ is introduced as, 
\begin{eqnarray}
  {\cal M}_{xy}^\pm=\left(
  \begin{array}{cc}
    {\cal M}_{xy} & 0 \\
    0 & \pm{\cal M}_{xy}^\ast
  \end{array}
  \right).
  \label{SM_eq8}
\end{eqnarray}
The sign $+$ ($-$) is adopted in the even-parity (odd-parity) superconducting state. 
Thus, we adopt ${\cal M}_{xy}^{+}$ in the BCS state while ${\cal M}_{xy}^{-}$ in the PDW state. 

\section{MIRROR CHERN NUMBER}

Because the BdG Hamiltonian ${\cal H}({\bm k})$ and the mirror operator ${\cal M}_{xy}^\pm$ are commutative,  
we can block-diagonalize the BdG Hamiltonian using the eigenbasis of mirror operator, 
\begin{eqnarray}
  V^\pm{\cal H}({\bm k})V^{\pm\dagger}=\left(
  \begin{array}{cc}
    {\cal H}_{\lambda=i}^\pm({\bm k}) & 0 \\
    0 & {\cal H}_{\lambda=-i}^\pm({\bm k})
  \end{array}
  \right).
  \label{SM_eq9}
\end{eqnarray}
The unitary matrix $V^\pm$ is obtained by the eigenbasis of ${\cal M}_{xy}^\pm$, and 
the subsector Hamiltonian is characterized by the eigenvalues, $\lambda=\pm i$. 
  Examples of subsector Hamiltonian for the bilayer system and the trilayer PDW state for $\lambda=i$ are given in Eqs.~(3) and (4). 
The subsector Hamiltonian for the BCS state in trilayers is given by
\begin{widetext}
  \begin{eqnarray}
    {\cal H}_{\lambda=i}^+({\bm k})=\left(
    \begin{array}{cccccc}
      \xi_\uparrow({\bm k}) & \alpha k_+ & \sqrt{2}t_\perp & 0 & d_{{\rm out}-}({\bm k}) & \psi_{\rm out} \\
      \alpha k_- & \xi_\downarrow({\bm k}) & 0 & 0 & -\psi_{\rm out} & -d_{{\rm out}+}({\bm k}) \\
      \sqrt{2}t_\perp & 0 & \xi_\uparrow({\bm k}) & \psi_{\rm in} & 0 & 0 \\
      0 & 0 & \psi_{\rm in}^\ast & -\xi_\downarrow({\bm k}) & 0 & -\sqrt{2}t_\perp \\
      d_{{\rm out}-}^\ast({\bm k}) & -\psi_{\rm out}^\ast & 0 & 0 & -\xi_\uparrow({\bm k}) & \alpha k_- \\
      \psi_{\rm out}^\ast & -d_{{\rm out}+}^\ast({\bm k}) & 0 & -\sqrt{2}t_\perp & \alpha k_+ & -\xi_\downarrow({\bm k})
    \end{array}
    \right),
    \label{SM_eq10}
  \end{eqnarray}
\end{widetext}
and
\begin{widetext}
  \begin{eqnarray}
    {\cal H}_{\lambda=-i}^+({\bm k})=\left(
    \begin{array}{cccccc}
      \xi_\uparrow({\bm k}) & \alpha k_+ & 0 & 0 & d_{{\rm out}-}({\bm k}) & \psi_{\rm out} \\
      \alpha k_- & \xi_\downarrow({\bm k}) & \sqrt{2}t_\perp & 0 & -\psi_{\rm out} & -d_{{\rm out}+}({\bm k}) \\
      0 & \sqrt{2}t_\perp & \xi_\downarrow({\bm k}) & -\psi_{\rm in} & 0 & 0 \\
      0 & 0 & -\psi_{\rm in}^\ast & -\xi_\uparrow({\bm k}) & -\sqrt{2}t_\perp  & 0 \\
      d_{{\rm out}-}^\ast({\bm k}) & -\psi_{\rm out}^\ast & 0 & -\sqrt{2}t_\perp & -\xi_\uparrow({\bm k}) & \alpha k_- \\
      \psi_{\rm out}^\ast & -d_{{\rm out}+}^\ast({\bm k}) & 0 & 0 & \alpha k_+ & -\xi_\downarrow({\bm k})
    \end{array}
    \right),
    \label{SM_eq11}
  \end{eqnarray}
\end{widetext}
while we obtain the $\lambda=-i$ subsector Hamiltonian for the PDW state as
\begin{widetext}
  \begin{eqnarray}
    {\cal H}_{\lambda=-i}^-({\bm k})=\left(
    \begin{array}{cccccc}
      \xi_\uparrow({\bm k}) & \alpha k_+ & 0 & 0 & -d_{{\rm out}-}({\bm k}) & -\psi_{\rm out} \\
      \alpha k_- & \xi_\downarrow({\bm k}) & \sqrt{2}t_\perp & 0 & \psi_{\rm out} & d_{{\rm out}+}({\bm k}) \\
      0 & \sqrt{2}t_\perp & \xi_\downarrow({\bm k}) & d_{{\rm in}+}({\bm k}) & 0 & 0 \\
      0 & 0 & d_{{\rm in}+}^\ast({\bm k}) & -\xi_\downarrow({\bm k}) & 0 & -\sqrt{2}t_\perp \\
      -d_{{\rm out}-}^\ast({\bm k}) & \psi_{\rm out}^\ast & 0 & 0 & -\xi_\uparrow({\bm k}) & \alpha k_- \\
      -\psi_{\rm out}^\ast & d_{{\rm out}+}^\ast({\bm k}) & 0 & -\sqrt{2}t_\perp & \alpha k_+ & -\xi_\downarrow({\bm k})
    \end{array}
    \right).
    \label{SM_eq12}
  \end{eqnarray}
\end{widetext}

The mirror Chern number $\nu(\lambda)$ is defined by
\begin{eqnarray}
  \nu(\lambda)&=&\frac{1}{2\pi}\int_{-\pi}^\pi\int_{-\pi}^\pi dk_xdk_y\epsilon^{ij}\partial_{k_i} A_{j,\lambda}^\pm({\bm k}), 
  \label{SM_eq13} \\
  A_{i,\lambda}^\pm({\bm k})&=&i\sum_{E_{n,\lambda}^\pm({\bm k})<0}\langle u_{n,\lambda}^\pm({\bm k})|\partial_{k_i}u_{n,\lambda}^\pm({\bm k})\rangle,
  \label{SM_eq14}
\end{eqnarray}
where $E_{n,\lambda}^\pm({\bm k})$ and $|u_{n,\lambda}^\pm({\bm k})\rangle$ are the eigenenergy and eigenstate 
of the subsector Hamiltonian ${\cal H}_\lambda^\pm({\bm k})$, namely, 
\begin{eqnarray}
  {\cal H}_\lambda^\pm({\bm k})|u_{n,\lambda}^\pm({\bm k})\rangle=E_{n,\lambda}^\pm({\bm k})|u_{n,\lambda}^\pm({\bm k})\rangle.
  \label{SM_eq15}
\end{eqnarray}
For the numerical calculation of the mirror Chern number, we adopt an efficient method developed 
in Ref.~\onlinecite{JPSJ.74.1674}.

\section{SYMMETRY CLASS OF THE SUBSECTOR HAMILTONIAN}

We examine the symmetry class of the subsector Hamiltonian ${\cal H}^\pm_\lambda({\bm k})$. 
For this purpose, we first consider the time-reversal symmetry, particle-hole symmetry, and chiral symmetry 
in the original BdG Hamiltonian, which are defined as, 
\begin{eqnarray}
  T{\cal H}({\bm k})T^\dagger&=&{\cal H}^\ast(-{\bm k}), 
  \label{SM_eq16} \\
  P{\cal H}({\bm k})P^\dagger&=&-{\cal H}^T(-{\bm k}), 
  \label{SM_eq17} \\
  C{\cal H}({\bm k})C^\dagger&=&-{\cal H}({\bm k}),
  \label{SM_eq18}
\end{eqnarray}
respectively. 
For bilayers, the operators $T$, $P$, and $C$ are given by 
\begin{eqnarray}
  T&=&{\rm diag}(i\sigma_y,i\sigma_y,i\sigma_y,i\sigma_y),
  \label{SM_eq19} \\
  P&=&\left(
  \begin{array}{cccc}
    0 & 0 & \sigma_0 & 0 \\
    0 & 0 & 0 & \sigma_0 \\
    \sigma_0 & 0 & 0 & 0 \\
    0 & \sigma_0 & 0 & 0
  \end{array}
  \right), 
  \label{SM_eq20} 
\end{eqnarray}
and $C=PT^\dagger $, respectively. 
As a result of the unitary transformation with use of $V^\pm$, Eqs.~(\ref{SM_eq16})-(\ref{SM_eq18}) 
are rewritten in terms of the subsector Hamiltonian. 
For the BCS state, we obtain  
\begin{eqnarray}
  &&\left(
  \begin{array}{cc}
    i\sigma_y & 0 \\
    0 & i\sigma_y
  \end{array}
  \right)
  {\cal H}_{\lambda=i}^+({\bm k})\left(
  \begin{array}{cc}
    -i\sigma_y & 0 \\
    0 & -i\sigma_y
  \end{array}
  \right)={\cal H}_{\lambda=-i}^{+\ast}(-{\bm k}), \nonumber \\
  \label{SM_eq21} \\
  &&\left(
  \begin{array}{cc}
    0 & \sigma_0 \\
    \sigma_0 & 0
  \end{array}
  \right){\cal H}_{\lambda=i}^+({\bm k})\left(
  \begin{array}{cc}
    0 & \sigma_0 \\
    \sigma_0 & 0
  \end{array}
  \right)=-{\cal H}_{\lambda=-i}^{+T}(-{\bm k}),
  \label{SM_eq22} \\
  &&\left(
  \begin{array}{cc}
    0 & -i\sigma_y \\
    -i\sigma_y & 0
  \end{array}
  \right){\cal H}_{\lambda}^+({\bm k})\left(
  \begin{array}{cc}
    0 & i\sigma_y \\
    i\sigma_y & 0
  \end{array}
  \right)=-{\cal H}_\lambda^+({\bm k}). 
  \label{SM_eq23}
\end{eqnarray}
Equation~(\ref{SM_eq21}) indicates that the time-reversal symmetry in the subsector is ill-defined unless 
${\cal H}_{\lambda=i}^+({\bm k})={\cal H}_{\lambda=-i}^+({\bm k})$, even when the time-reversal symmetry is conserved in 
the original BdG Hamiltonian. 
Only when the special condition ${\cal H}_{\lambda=i}^+({\bm k})={\cal H}_{\lambda=-i}^+({\bm k})$ is satisfied and 
Eq.~(\ref{SM_eq21}) holds, we can rely on the time-reversal symmetry in the subsector Hamiltonian. 
Similarly, the condition ${\cal H}_{\lambda=i}^+({\bm k})={\cal H}_{\lambda=-i}^+({\bm k})$ 
as well as Eq.~(\ref{SM_eq22}) have to be satisfied for the particle-hole symmetry in the subsector Hamiltonian. 
On the other hand, the chiral symmetry is well-defined in the subsector Hamiltonian as Eq.~(\ref{SM_eq23}). 

\begin{table*}[tbp]
  \begin{tabular}{c||c|c|c|c}
    \hline\hline
    & $t_\perp=0$, $H=0$ & $t_\perp\neq 0$, $H=0$ & $t_\perp=0$, $H\neq 0$  & otherwise \\
    \hline\hline
    BCS state & DIII  & AIII  & D & A \\
    \hline
    PDW state & DIII  & D & D & D \\
    \hline\hline
  \end{tabular}
  \caption{Symmetry class of the subsector Hamiltonian ${\cal H}_\lambda^\pm({\bm k})$ for the bilayer SC.}
  \label{SM_tab1}
\end{table*}
In the same way, the time-reversal symmetry and the particle-hole symmetry in the PDW state are described with use 
of the subsector Hamiltonian as, 
\begin{eqnarray}
  &&\left(
  \begin{array}{cc}
    i\sigma_y & 0 \\
    0 & i\sigma_y
  \end{array}
  \right)
  {\cal H}_{\lambda=i}^-({\bm k})\left(
  \begin{array}{cc}
    -i\sigma_y & 0 \\
    0 & -i\sigma_y
  \end{array}
  \right)={\cal H}_{\lambda=-i}^{-\ast}(-{\bm k}), \nonumber \\
  \label{SM_eq24} \\
  &&\left(
  \begin{array}{cc}
    0 & \sigma_0 \\
    \sigma_0 & 0
  \end{array}
  \right){\cal H}_{\lambda}^-({\bm k})\left(
  \begin{array}{cc}
    0 & \sigma_0 \\
    \sigma_0 & 0
  \end{array}
  \right)=-{\cal H}_{\lambda}^{-T}(-{\bm k}), 
  \label{SM_eq25} 
\end{eqnarray}
respectively. 
According to Eq.~(\ref{SM_eq24}), the particle-hole symmetry is well-defined in the subsector Hamiltonian, 
and it is always conserved in the PDW state. Thus, we do not have to consider the chiral symmetry because it 
coincides with the time-reversal symmetry. The time-reversal symmetry is conserved in the subsector Hamiltonian 
only when the special condition ${\cal H}_{\lambda=i}^-({\bm k})={\cal H}_{\lambda=-i}^-({\bm k})$ is satisfied and 
Eq.~(\ref{SM_eq25}) holds.
In Table \ref{SM_tab1}, we summarize the symmetry class of the subsector Hamiltonian for the bilayer BCS and PDW states.

The symmetry class of the subsector Hamiltonian in the trilayer SC can be analyzed in the same way. 
For the BCS state, the time-reversal symmetry, particle-hole symmetry, and chiral symmetry are given by
\begin{widetext}
  \begin{eqnarray}
    &&\left(
    \begin{array}{ccc}
      -i\sigma_y & 0 & 0 \\
      0 & \sigma_z & 0 \\
      0 & 0 & -i\sigma_y
    \end{array}
    \right){\cal H}_{\lambda=i}^+
    \left(
    \begin{array}{ccc}
      i\sigma_y & 0 & 0 \\
      0 & \sigma_z & 0 \\
      0 & 0 & i\sigma_y
    \end{array}
    \right)={\cal H}_{\lambda=-i}^{+\ast}(-{\bm k}),
    \label{SM_eq26}\\
    &&\left(
    \begin{array}{ccc}
      0 & 0 & \sigma_0 \\
      0 & \sigma_x & 0 \\
      \sigma_0 & 0 & 0
    \end{array}
    \right){\cal H}_{\lambda=i}^+({\bm k})
    \left(
    \begin{array}{ccc}
      0 & 0 & \sigma_0 \\
      0 & \sigma_x & 0 \\
      \sigma_0 & 0 & 0 
    \end{array}
    \right)=-{\cal H}_{\lambda=-i}^{+T}(-{\bm k})
    \label{SM_eq27}\\
    &&\left(
    \begin{array}{ccc}
      0 & 0 & -i\sigma_y \\
      0 & \mp i\sigma_y & 0 \\
      -i\sigma_y & 0 & 0 
    \end{array}
    \right){\cal H}_{\lambda=\pm i}^+({\bm k})
    \left(
    \begin{array}{ccc}
      0 & 0 & i\sigma_y \\
      0 & \pm i\sigma_y & 0 \\
      i\sigma_y & 0 & 0
    \end{array}
    \right)=-{\cal H}_{\lambda=\pm i}^+({\bm k}),
    \label{SM_eq28}
  \end{eqnarray}
\end{widetext}
respectively. 
On the other hand, the time-reversal symmetry and particle-hole symmetry in the PDW state are given by   
\begin{eqnarray}
  &&\left(
  \begin{array}{ccc}
    -i\sigma_y & 0 & 0 \\
    0 & \sigma_0 & 0 \\
    0 & 0 & -i\sigma_y
  \end{array}
  \right){\cal H}_{\lambda=i}^-
  \left(
  \begin{array}{ccc}
    i\sigma_y & 0 & 0 \\
    0 & \sigma_0 & 0 \\
    0 & 0 & i\sigma_y
  \end{array}
  \right)={\cal H}_{\lambda=-i}^{-\ast}(-{\bm k}), \nonumber \\
  \label{SM_eq29}\\
  &&\left(
  \begin{array}{ccc}
    0 & 0 & \sigma_0 \\
    0 & \sigma_x & 0 \\
    \sigma_0 & 0 & 0
  \end{array}
  \right){\cal H}_{\lambda}^-({\bm k})
  \left(
  \begin{array}{ccc}
    0 & 0 & \sigma_0 \\
    0 & \sigma_x & 0 \\
    \sigma_0 & 0 & 0 
  \end{array}
  \right)=-{\cal H}_{\lambda}^{-T}(-{\bm k}). \nonumber \\
  \label{SM_eq30}
\end{eqnarray}
The particle-hole symmetry is always conserved in the subsector Hamiltonian of the PDW state. 
We summarize the symmetry class of the subsector Hamiltonian for the BCS and PDW states  
in Tables \ref{SM_tab2} and \ref{SM_tab3}, respectively.
\begin{table}[tbp]
  \begin{tabular}{c|c|c|c}
    \hline\hline
    & $t_\perp=0$, $H=0$ & $t_\perp\neq 0$, $H=0$ &  otherwise \\
    \hline
    Symmetry class & DIII & AIII  & A  \\
    \hline\hline
  \end{tabular}
  \caption{Symmetry class of the subsector Hamiltonian ${\cal H}_\lambda^+({\bm k})$ in the trilayer BCS state.}
  \label{SM_tab2}
\end{table}
\begin{table}[tbp]
  \begin{tabular}{c|c|c}
    \hline\hline
    & $t_\perp=0$, $H=0$, ${\bm d}_{\rm in}=0$ & otherwise \\
    \hline
    Symmetry class & DIII  & D \\
    \hline\hline
  \end{tabular}
  \caption{Symmetry class of the subsector Hamiltonian ${\cal H}_\lambda^-({\bm k})$ in the trilayer PDW state.}
  \label{SM_tab3}
\end{table}

It is straightforward to elucidate the symmetry class of more than three layers. 
Independent of the number of layers, the subsector Hamiltonian belongs to the class D (class A) 
in the PDW state (BCS state), under the realistic conditions $t_\perp \ne 0$ and $H \ne 0$.

\section{TOPOLOGICAL NUMBER OF SPINLESS CHIRAL P-WAVE SC}
The topological number of spinless chiral $p$-wave SC part in Eqs.~(\ref{eq4}) and (\ref{SM_eq12}) 
is given by~\cite{JPSJ.81.011013}
\begin{eqnarray}
\hspace{-6mm}
\nu_{\rm p}(\lambda)=-\frac{1}{2}\sum_{\eta_\lambda({\bm k}^0)=0}{\rm sgn}[\xi_\lambda({\bm k}^0)]{\rm sgn}[{\rm det}\partial_{k_i}\eta_{\lambda,j}({\bm k}^0)],
\label{SM_eq31}
\end{eqnarray}
where $\xi_\lambda({\bm k})=\xi_\uparrow({\bm k})$  and $\eta_\lambda({\bm k})=-d_{\rm in-}({\bm k})$  
for $\lambda=i$, and 
$\xi_\lambda({\bm k})=\xi_\downarrow({\bm k})$ and $\eta_\lambda({\bm k})=d_{\rm in+}({\bm k})$ for $\lambda=-i$. 
We denoted $ \eta_\lambda({\bm k})  = \eta_{\lambda,1}({\bm k}) +i  \eta_{\lambda,2}({\bm k})$.
We find $\nu_{\rm p}(\lambda)$ to be non-zero if the odd number of zero-nodes of $ \eta_{\lambda}({\bm k}) $ [$\eta_\lambda({\bm k}^0)=0$] 
are enclosed by the Fermi surface. 
Indeed, one zero node is enclosed by the Fermi surface for the simple dispersion adopted in this paper.
Thus, we obtained the nontrivial Chern number $\nu_{\rm p}(\pm i) = \mp 1$,
which is identified as the mirror Chern number of the subsector Hamiltonian Eqs.~(\ref{eq4}) and (\ref{SM_eq12}) in the limit $t_{\perp} \rightarrow 0$.
}
  
%\bibliography{reference}
\input{paper.bbl}
\end{document}